\documentclass[doublecol]{epl2}

 \usepackage{graphicx}
 \RequirePackage{color}

 \definecolor{MyDarkGreen}{rgb}{0.02,0.60,0.06}

\title{Universal Properties of Mythological Networks}
\shorttitle{Universal Properties of Mythological Networks} 

\author{P{\'{a}}draig Mac~Carron \and Ralph Kenna}

\institute{
  Applied Mathematics Research Centre, Coventry University, Coventry, CV1 5FB, England, EU
}

\abstract{As in statistical physics, the concept of universality plays an important, albeit qualitative, role in the field of comparative mythology.
Here we apply statistical mechanical tools to analyse the networks underlying three iconic mythological narratives with a view to identifying common and distinguishing quantitative features.
Of the three narratives, an Anglo-Saxon and a Greek text are mostly believed by  antiquarians to be partly historically based while the third, an Irish epic, is often considered to be fictional.
Here we  use network analysis  in an attempt to discriminate real from imaginary social networks and place mythological narratives
on the spectrum between them.
 This suggests that the perceived artificiality of the  Irish narrative can be traced back to anomalous features associated with six characters.
 Speculating that these are amalgams of  several entities or proxies, renders the plausibility of the Irish text comparable to the others 
from a network-theoretic point of view.
}
\begin{document}

\maketitle

\section{Introduction}


Over the past decades many statistical physicists have turned their attention to other disciplines
in attempts to understand how properties of complex systems emerge from the interactions
between component parts in a non-trivial manner.
Applications include the analysis of complex networks in the natural, social and technological sciences as well as in the humanities
\cite{Costa2009,AlbertBarabasi,Ne03,Wasserman,Stauffer}.
One of the notions intrinsic to statistical physics is universality, and attempts have been made to
classify complex networks from a variety of areas to facilitate comparison amongst them
\cite{Amaral,Onnela}.

Universality is also an important, albeit hitherto qualitative,  notion in the field of comparative mythology.
Campbell maintained that mythological narratives from a variety of cultures  essentially share the same fundamental structure, called the \emph{monomyth}~\cite{Campbell}.
Here we statistically
compare networks underlying mythological narratives from three different cultures to each other as well as to real, imaginary and random networks. In this way we  quantitatively explore universality in mythology and attempt to place mythological narratives on the spectrum from the real to the imaginary.


Network theory has recently been developed and applied to polymers, economics, particle physics, computer science, sociology, biology, epidemiology, linguistics and more \cite{Costa2009,WattsStrogatz,Wasserman,Stauffer,AlbertBarabasi,Amaral,Ne03,Onnela,Yurko,Liljeros}.
\begin{table*}
 \caption{Size, mean degree, mean path length, diameter, clustering coefficient,  absolute (relative) size of giant component  and assortativity $r$ for the  epics, together with mean path length $\ell_{\rm{rand}}$ and clustering $C_{\rm{rand}}$ for similarly sized random networks.}
    \vspace{0.2cm}
    \centering
  \begin{tabular}{l|c c c c c c c c c c c}
          Network & & $N$  &    $\langle{k}\rangle$    & $\ell$ & $\ell_{\rm{rand}}$ & $\ell_{\rm{max}}$& $C$ & $C_{\rm{rand}}$ & 
           $G_c$ &  $r$\\
    \hline
  \textbf{Beowulf}   & All      & ~74 & 4.45 & 2.37 & 2.88 & ~6 & 0.69  & 0.06  &~50 (67.5\%)  & -0.10\\
                     & Hostile  & ~31 & 1.67 & 2.08 & 3.25 & ~4 & 0~~~~ & 0.05  &  ~10 (32.2\%)  & -0.20\\
                     & Friendly & ~68 & 4.12 & 2.45 & 2.98 & ~6 & 0.69  & 0.06  & ~45 (66.1\%)  & -0.03\\
    \hline
  \textbf{T\'ain}    & All	    & 404	& 6.10 & 2.76 & 3.32 & ~7 & 0.82  & 0.02  & 398 (98.5\%)  & -0.33 \\
                     & Hostile  & 144	& 2.33 & 2.93 & 5.88 & ~7 & 0.17  & 0.02  & 131 (90.9\%)  & -0.36\\
                     & Friendly & 385	& 5.67 & 2.84 & 3.43 & ~7 & 0.84  & 0.01  & 350 (90.9\%)  & -0.32\\
    \hline
   \textbf{Iliad}    & All	    & 716	& 7.40 & 3.54 & 3.28 & 11 & 0.57  & 0.01   & 707 (98.7\%)  & -0.08 \\
                     & Hostile  & 321	& 2.25 & 4.10 & 7.12 & ~9 & 0~~~~ & 0.01   & 288 (89.4\%)  & -0.39 \\
                     & Friendly & 664	& 6.98 & 3.83 & 3.34 & 12 & 0.62  & 0.01   & 547 (82.3\%)  & ~0.10 \\
                     \hline
  \textbf{Beowulf*}  & Friendly & ~67 & 3.49 & 2.83 & 3.36 & ~7 & 0.68  & 0.05  & ~43 (64.2\%)  & ~0.01\\
  \textbf{T\'ain*}   & Friendly & 324	& 3.71 & 3.88 & 4.41 & ~8 & 0.69  & 0.01 & 201 (62.0\%)  & ~0.04\\
 \end{tabular}
 \label{tab:data}
\end{table*}
A number of statistics have been developed to capture features of such networks.
Some  structural properties are quantified using the characteristic path length $\ell$, the longest geodesic $\ell_{\rm{max}}$ and the clustering coefficient $C$.
The first of these  is a measure of the average minimum separation between pairs of $N$ nodes and $\ell_{\rm{max}}$ is the diameter of the network.
Clustering measures to what extent a given neighbourhood of the network is cliqued.
If node $i$ has $k_i$ neighbours, then the maximal number of potential edges between them is $k_i(k_i-1)/2$.
If $n_i$ is the actual number of bonds between the $k_i$ neighbours of $i$, the clustering coefficient of the node is \cite{WattsStrogatz}
\begin{equation}
 C_i = \frac{2n_i}{k_i(k_i-1)}.
\label{ci}
 \end{equation}
Many complex networks have a modular structure implying that groups of nodes organise in a hierarchical manner into increasingly larger groups. Hierarchical networks are characterised by a power-law dependency of the clustering coefficient on the node  degree \cite{Ravasz1,Dorogovtsev,Gleiser},
\begin{equation}
C(k) \sim \frac{1}{k}.
\label{Cvsk}
\end{equation}
The mean degree $\langle{k}\rangle$ is obtained by averaging over the nodes of the network  and its clustering coefficient $C$ is obtained by averaging eq.(\ref{ci}).
A network is said to be small world if $\ell \approx \ell_{\rm{rand}}$  and $C \gg C_{\rm{rand}}$, where $\ell_{\rm{rand}}$ is the average path length and $C_{\rm{rand}}$ is the clustering coefficient of a random network of the same size and average degree~\cite{WattsStrogatz}.

If $p(k)$ is the probability that a given vertex has degree $k$, then the degree distribution for many networks is
\begin{equation}
 p(k) \sim k^{-\gamma},
 \label{powerlaw}
\end{equation}
for positive constant $\gamma$, with perhaps a cut-off for high degree.
Such power laws indicate that, while most nodes are sparsely connected, some are linked to many others and play an especially important role in maintaining network integrity.
Networks are scale free if the power law (\ref{powerlaw}) holds with $2 < \gamma \le 3$ \cite{Amaral}.
Assortative mixing by degree is the notion that vertices of high degree associate with similarly highly connected vertices, while vertices of low degree associate with other less linked nodes.

Another measure of the connectivity of the network is  the size of the giant component $G_c$. In a scale-free network, removal of influential nodes causes the giant component to break down quickly whereas it remains intact upon randomly removing nodes from the network \cite{Albert2}. As well as the degree, the betweenness centrality of a node $g_l$ indicates how influential that node is, it controls the flow of information between other vertices. It is the total number of geodesics (shortest paths) that pass through a given node \cite{Freeman}.
If $\sigma(i, j)$ is the number of geodesics between nodes $i$ and $j$,
and if the number of these which pass through node $l$ is $\sigma_l(i, j)$, the betweenness centrality of  vertex $l$ is
\begin{equation}
g_l = \frac{2}{(N-1)(N-2)}\sum_{i\neq j} \frac{\sigma_l(i,j)}{\sigma(i,j)}.
\end{equation}
The normalization ensures that $g_l=1$ if all geodesics pass through $l$.

The above statistics capture a variety of characteristics of  networks and, by comparing them  between different networks, we get an idea how similar or different they are.
These quantitative indicators therefore play similar roles to critical exponents in the study of phase transitions.


Social networks are usually small world \cite{WattsStrogatz,Amaral},  highly clustered, assortatively mixed by degree \cite{Newman5,Newman6} and scale free \cite{Amaral,Barabasi2}.
Some real social networks that have been studied include between company directors \cite{Davis}, jazz musicians \cite{GleiserJazz}, movie actors \cite{WattsStrogatz,Amaral}, scientific co-authors \cite{Newman1,Newman2,Newman3,Barabasi1} as well as online social networks  \cite{Leskovec,Szell,Facebook}.
Catalogues of the properties of these and other networks are contained in, \emph{e.g.},  refs.\cite{AlbertBarabasi,Ne03}.
In refs.\cite{Alberich, Gleiser} the social network of characters appearing in Marvel comics (the so-called \emph{Marvel Universe}) was constructed. In that network two characters are connected if they appear in the same comic book. This is clearly an artificial social network as the characters in this universe represent an imaginary society, but it is a social network nonetheless. Analysis showed that, while it mimicked other social networks to an extent, it was unable to hide its artificial nature~\cite{Alberich,Gleiser}.
To facilitate comparison between our mythological networks and other real and imaginary networks, we also look here at four works of fiction.
With these at our disposal, we seek to compare mythological narratives to other networks, ranging from the real to the imaginary.

\section{Comparative mythology}


The mythological narratives studied here are \emph{Beowulf} \cite{Beowulf}, the \emph{Iliad} \cite{Iliad} and the \emph{T\'ain B\'o Cuailnge} \cite{Tain}.
In comparative mythology, these have  statuses similar to that of the Ising model in statistical physics.
They are  widely studied, frequently compared to each other  and still present puzzles which continue to be investigated.
\emph{Beowulf} is an Old English heroic epic, set in Scandinavia.
A single codex survives which is dated from between the 8th and early 11th centuries \cite{Benson,Beowulf2}.
Although the poem is embellished by obvious fiction, archaeological excavations in Denmark and Sweden support historicity associated with some of the human characters although the main character Beowulf is mostly not believed to have existed \cite{Chambers,Klaeber}.
The \emph{Iliad} is an epic poem attributed to Homer and is dated to the 8th century BC \cite{Iliad}.
Recent evidence suggests that the story may be based on a historical conflict during the 12th century BC~\cite{MKorf,Kraftetc}.
The \emph{T\'ain B\'o Cuailnge} (\emph{``T\'ain''} from here on) is an Irish epic, surviving in three 12th and 14th century manuscripts.
It describes a conflict between  Connaught and Ulster, Ireland's western and northern provinces.
Before it was committed to writing, the {\emph{T\'ain}} had an extensive oral history. It was dated by medieval  scholars to the first centuries BC, but this may have been an attempt by Christian monks to artificially synchronise oral traditions with biblical and classical history \cite{Tain}.
Its historicity is often questioned.
Jackson (1964) argues that such narratives corroborate Greek and Roman accounts of the Celts and offer us a `window on the iron age' \cite{Jackson} but O'Rahilly (1964) objects that such tales have no historical basis whatsoever \cite{Rahilly}.


From the databases created for each mythological epic, 74 unique characters were identified in \emph{Beowulf}, 404 for the \emph{T\'ain} and 716 for the \emph{Iliad}.
(The reader should keep in mind that these networks, like many others \cite{AlbertBarabasi}, are necessarily of limited extent and incomplete - potentially representing spotlights on the societies from which they are drawn.)
We define two distinct relationship types: \emph{friendly}  and  \emph{hostile} edges. Friendly links are made if two characters are related; speak directly to one another; speak about one another or are present together and it is clear they know each other. Hostile links are made when two characters meet in a conflict and a friendly link is not made if they speak here and only here; or when a character explicitly declares animosity against another  and it is clear they know each other.
To ensure consistency throughout, we first separately constructed the network for one of the narratives.
Comparing our individual interpretations allowed us to tune the rules to a consistent and implementable set and to satisfy ourselves that the networks could be constructed in a reasonably objective manner.
We also inspected different translations of the original texts to verify that these did not result in significant differences to our quantitative results.


In table~\ref{tab:data} a list of statistics for the networks underlying each epic is compiled.
These include
the mean degree  $\langle{k}\rangle$, the average path length $\ell$, diameter $\ell_{\rm{max}}$, clustering coefficient $C$, the
size of the giant component $G_c$ in absolute terms and as a percentage of the size $N$, and the assortativity $r$.
In addition the average path length $\ell_{\rm{rand}}$   and the clustering coefficient $C_{\rm{rand}}$
for a random network of the same size and average degree are listed.
The statistics indicate that the complete networks are similar to the friendly ones reflecting that, even though conflict is an element of each narrative, they are still stories about human relations, and the discussions rather than disputes drive the stories.
Like real social networks the mythological networks have
average path lengths similar to those of random networks of the same size and average degree.
They also have high clustering coefficients compared to random networks indicating they are small world.
The hostile networks are quite different; the facts that their average path lengths are smaller than
those of random networks and that they have virtually no clustering tell us they are not small world.
The reason for such low clustering in hostile networks relates to the idea of `the enemy of my enemy is my friend' --
in general a common foe tends to suppress hostilities between two nodes.


In the overall network, closed triads with just one hostile edge are also disfavoured as a single hostile link prompts the opposite node in a triangle to take sides. The propensity to disfavour odd numbers of hostile links in a closed triad is known as structural balance \cite{Heider,Cartwright}. Structural balance  has been observed in real-world situations, such as the shifting alliances in the lead-up to war \cite{Antal}.
We find similar results in our three networks with only 3.8\% of {\emph{Beowulf's}} closed triads containing an odd number of hostile links,
8.0\% of the {\emph{T\'ain's}} and just  1.7\% of the  {\emph{Iliad's.}}

\begin{figure}[t]
\begin{center}
\includegraphics[width=0.6\columnwidth, angle=0]{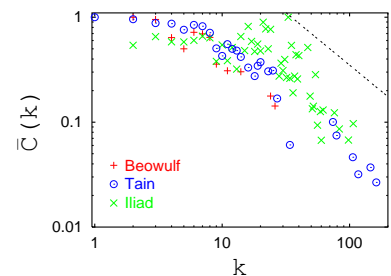}
\caption{Mean clustering coefficient per degree $\bar{C}(k)$ vs $k$ for the three myths with the  power law $1/k$ as a dashed line   to guide the eye.
}\label{fig1}
\end{center}
\end{figure}

In fig.\,\ref{fig1} we test for hierarchical structure by plotting the mean clustering coefficient per degree $\bar{C}(k)$ versus $k$ and comparing with eq.(\ref{Cvsk}).
The nodes with a smaller degree present higher clustering than those with larger degree and the decay approximately follows eq.(\ref{Cvsk}).
The nodes with low degree are part of densely interlinked clusters, while the vertices with high degree bring together the many small communities of clusters into a single, integrated network.

In most collaboration networks, a very large subset of nodes are connected to each other (the giant component).
The full \emph{T\'ain} and \emph{Iliad} networks contain about 98\% of the nodes.
\emph{Beowulf} has a smaller giant component (68\% of all nodes) because the narrative 
contains two stories dealing with events from the past, disconnected from the main plot.
Usually for collaboration networks, the giant components contain
 less than 90\% of all nodes \cite{Newman2} and the friendly networks of each epic falls into this category. 
Removing the top 5\% of nodes with the highest betweenness centrality causes  all three networks to break down quickly,
reducing the giant component of \emph{Beowulf} and the \emph{T\'ain} by over 30\%.
This lack of robustness shows how reliant they are on the most connected characters.
If, however, 5\% of the nodes are removed at random the giant component in each case is largely unaffected.
Vulnerability to targeted attack but robustness to random attack, and the hierarchical structure of the networks, hint that these networks may be scale free.

\begin{figure*}[t]
\begin{center}
\includegraphics[width=0.6\columnwidth, angle=0]{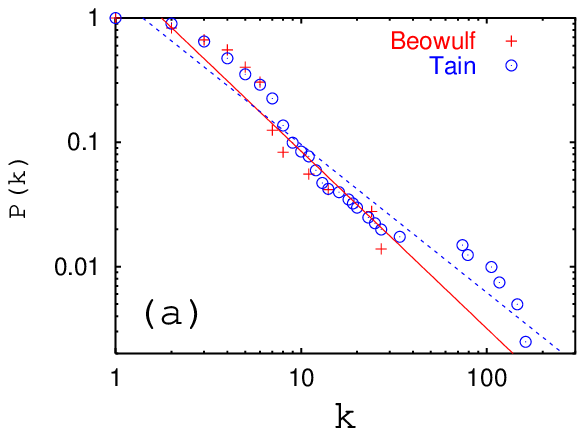}
\includegraphics[width=0.6\columnwidth, angle=0]{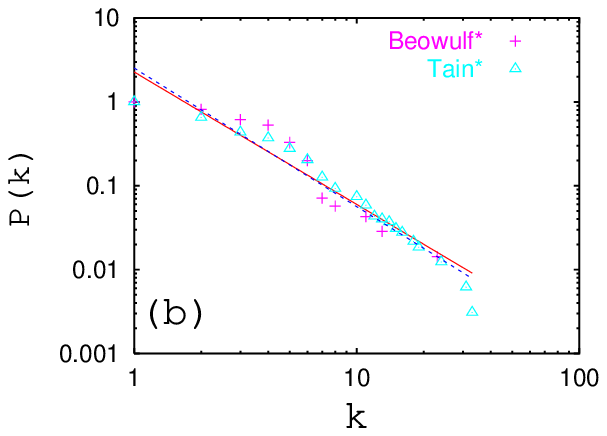}
\includegraphics[width=0.6\columnwidth, angle=0]{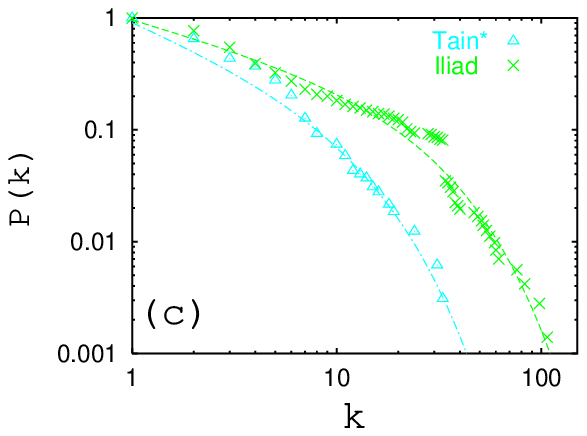}
\caption{The degree distribution (a) as a power law for \emph{Beowulf} and the \emph{T\'ain}
(b) for \emph{Beowulf*} and the \emph{T\'ain*}  and (c) as truncated power laws for the \emph{T\'ain*} and the  \emph{Iliad}.}
\label{fig2}
\end{center}
\end{figure*}



Newman showed that real social networks tend to be assortatively mixed by degree \cite{Newman5}
and Gleiser demonstrated that disassortativity of social networks may signal artificiality \cite{Gleiser}.
However disassortivity may also reflect the conflictual nature of the stories.
In the \emph{T\'ain} and the \emph{Iliad}, in particular, characters are frequently introduced to fight one of the heroes and are killed virtually immediately.
These encounters link the high-degree heroes to low-degree victims and may explain why the hostile networks are highly disassortative.
This suggests that only in the friendly networks may disassortativity be confidently interpreted as signaling artificiality.

The assortativity coefficient $r$ is given by the Pearson correlation coefficient of the degrees between all pairs of linked nodes and is listed for each network in table~\ref{tab:data}.
Positive correlation indicates assortative mixing and a negative value indicates disassortativity.
As expected, all hostile networks are strongly disassortative.
However, the \emph{Iliad} friendly network is  assortative and the \emph{Beowulf} friendly network is only mildly disassortative.
\emph{Beowulf} takes place in two different settings, years apart, with the eponymous protagonist being the only character to appear in both places.
As it is a small network, the main character has high degree but most of his neighbours have lower degree rendering the network disassortative.
Indeed, removing the protagonist (labeling the resulting network {\emph{Beowulf*}}) delivers a positive assortativity coefficient ($r=0.012$) for the interactions between the remaining characters.

Thus  the \emph{Iliad} and {\emph{Beowulf*}} friendly networks have properties associated with real social networks. While the {\emph{T\'ain}} has many such features, it is not assortative.
This corroborates antiquarians' interpretations of the historicity of these myths (obviously fabulous entities and interactions notwithstanding)
-- the societies in the \emph{Iliad} and  \emph{Beowulf*} (without the eponymous protagonist) may be based on reality while that of  the {\emph{T\'ain}} appears fictional \cite{Rahilly}.
We next turn our attention to
networks which are definitely fictional, before
an analysis of degree distributions which will illuminate the reasons behind
the apparent artificiality exhibited by the \emph{T\'ain}.

\section{Fictional Narratives}

The above characteristics of the three mythological networks distinguish them from the intentionally fictitious Marvel universe studied in refs.\cite{Alberich, Gleiser}.
The question arises whether these characteristics are properties of non-comic fictional literature in general or whether this may truly signal a degree of historicity underlying the three  mythological narratives.
To investigate this, we applied our network tools to four  narratives from fictional literature.
These are Hugo's \emph{Les Mis\'erables} ($N=77$), 
 Shakespeare's \emph{Richard III} ($N=70$),  
 Tolkien's  \emph{Fellowship of the Ring} (the first part of the \emph{Lord of the Rings} trilogy, with $N=118$) 
 and Rowling's \emph{Harry Potter} ($N=72$).
We found that each of these, together with the {\emph{Marvel}} universe ($N=6\,846$),  has a very high clustering coefficient, is disassortative and has a giant component containing almost every character.

Attacking the smaller fictional networks by removing the top five characters with the highest betweenness centralities, we find that the networks are quite robust. The giant component goes from 100\% to 73.6\% for \emph{Les Mis\'erables}, 94.3\% to 84.6\% for \emph{Richard III}, 94.1\% to 85.8\% for \emph{Fellowship of the Ring} and 97.2\% to 77.4\% for \emph{Harry Potter}. Removing the top 10\% of nodes with highest betweenness shows \emph{Les Mis\'erables} and \emph{Harry Potter} are less robust than the others, however the Marvel Universe is barely affected by losing the top 684 vertices. This robustness is also indicative of an exponential degree distribution (see below and ref.~\cite{Albert2}). The robustness and the high clustering coefficient shows how well connected the networks are.
All five fictional networks have a hierarchical structure.
While these networks display the high clustering coefficient that is common to all social networks, the fact that they are all disassortative and are almost entirely connected
is perhaps an indication of their societies' artificiality. In a sense they are too small world to be real.

\section{Degree distributions}

We finally turn to the degree distributions associated with the various networks.
This will allow us to identify the source of the disassortativity of the \emph{T\'ain}  in particular and to speculate as to what it would take to render that network more realistic.


The cumulative distribution functions for the three narratives are plotted in fig.\,\ref{fig2}.
Estimates from eq.(\ref{powerlaw}) for \emph{Beowulf}, the \emph{T\'ain} and the \emph{Iliad} yield $\gamma = 2.4 \pm 0.2, 2.2 \pm 0.1$ and $2.4 \pm 0.1$ (with $\chi^2/$df 0.3, 0.2 and 0.4), respectively.
The degree exponents for the friendly networks are very similar to these values.
These results signal that the degree distributions of these networks are indeed scale free.

For comparison, the degree distribution for the fictional narratives only follows a power law in the tail in one instance (\emph{Harry Potter}) and only this appears to be scale free. The other three, as the Marvel Universe, are better described by exponential distributions.

In fig.\,2(a) we observe a striking similarity between the degree distributions of \emph{Beowulf} and the \emph{T\'ain} for all but the six most connected characters in the latter.
In fact, omitting the corresponding 6 points delivers the estimate $\gamma = 2.4 \pm 0.1$, the same value as for \emph{Beowulf} and the \emph{Iliad}.
However, we have no legitimate basis to omit the six most important characters of the narrative, especially if the degree distribution is scale free, because our previous results indicate this would destroy the giant component.
Instead we speculate that these six characters are in fact amalgams of several entities or proxies, whose collective degrees are large, but whose individual degrees are reduced.
To test this hypothesis, we removed the weak links (where they interact with characters only once in the entire narrative) from the characters
and denote the result by \emph{T\'ain*}.
In fig.\,2(b),  this is  compared to \emph{Beowulf*} as a power law.
Because six characters now have lower degree, the  estimate for $\gamma$ increases.
The modified exponent is $\gamma = 2.65 \pm 0.10$ ($\chi^2/{\rm{df}} = 0.15$), close to
for {\emph{Beowulf*}}, which has  $\gamma = 2.58 \pm 0.19$ ($\chi^2/{\rm{df}} = 0.27$).
The steeper slope, particularly for large $k$, is indicative of the presence of a fast decay and is common in many networks.
To understand this a comparison of the  \emph{T\'ain} with the larger network of the {\emph{Iliad}}  is appropriate.

The \emph{Iliad} is, in fact, better fitted by a truncated power law $P(k) \sim k^{1-\tau}\exp{(- {k}/{k^*}})$ with $\tau = 1.51 \pm 0.03$ ($\chi^2/{\rm{df}} = 0.07$).
The  \emph{T\'ain*} is compared to \emph{Iliad} in fig.\,2(c). It is also well fitted by
a truncated power law and delivers  $\tau = 1.74 \pm 0.06$  ($\chi^2/{\rm{df}}  = 0.03$).

The modified friendly \emph{T\'ain*} network has assortativity coefficient $r = 0.042$.
This positive value, together with the fact that the fitted degree distribution captures the top six data,
ensures that the networks share all the properties of real social networks.
(In comparison, simply removing the top six characters results in $ r = -0.122$ and destroys the giant component.)

Another way to test for assortativity is to plot the degree of the neighbours of a vertex as a function of its degree.
Positive slope indicates assortativity and negative slope signals disassortativity.
Plots for the averages of the neighbouring degrees are contained in fig.\,3.
{\emph{Beowulf}} and the {\emph{Iliad}} are assortative, while  the \emph{T\'ain} is clearly disassortative.
However, removing weak links associated with the six strongest characters  (resulting in the \emph{T\'ain*} network) again renders that Irish narrative assortative,   suggesting possibly historicity on a level comparable to {\emph{Beowulf}}.


\begin{figure}[t]
\begin{center}
\includegraphics[width=0.6\columnwidth, angle=0]{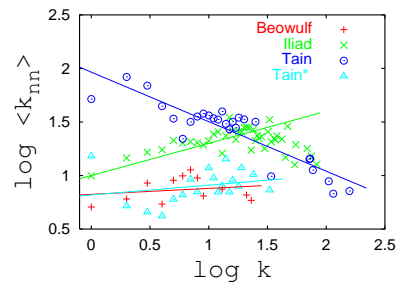}
\caption{The average $\langle{k_{\rm{nn}}}\rangle$ of the degrees of the neighbours of  vertices of degree $k$ and accompanying trend lines.
Positive (negative) trends indicate assortativity (disassortativity).}
\label{fig3}
\end{center}
\end{figure}

\section{Conclusions}

We have analysed the networks of relationships between characters of three mythological epics and four fictional
narratives and compared them to each other and to other social networks, both real and imaginary.
We found that all seven networks are small world, highly clustered, hierarchical and resilient to random attack --
properties generally associated with real social networks.
The fictional networks are, however, discernable from real social networks in that they mostly have exponential degree distributions, have relatively larger giant components, are robust under targeted attack and are dissassortative.
Moreover, they share this latter set of properties with the network underlying the 
Marvel universe previously studied in refs.\cite{Alberich,Gleiser}.

In an attempt to place the three mythological networks on the spectrum from the real to the fictitious, we compared their properties to actual and imaginary social networks.
Table~\ref{tableprops} summarises the broad properties of the different types of networks.
Of the three myths, the network of characters in the \emph{Iliad} has  properties most similar to those of real social networks.
It has a power-law degree distribution (with an exponential cut-off), is small world, assortative,
vulnerable to targeted attack and is structurally balanced.
This similarity perhaps reflects the  archaeological evidence supporting the historicity of some of the events of the \emph{Iliad}.

There is also archaeological evidence suggesting some of the characters in \emph{Beowulf} are based on real people, although the events in the story often contain elements of fantasy associated with the eponymous protagonist. The network for this society, while small, has some properties similar to real social networks, though like all the fictional narratives it is disassortative.
However, removing the main character from the network renders it assortative.
Thus, while the entire network is not credible as reflecting a real society, we suggest that
an assortative subset has properties akin to real social networks, and this subset has corroborative evidence of historicity.

\begin{table}
\caption{Summary of properties of mythological networks (\emph{Beowulf*}, \emph{T\'{a}in*} and \emph{Iliad}) compared to social and fictional networks. Here, small world implies $\ell \approx \ell_{\rm{rand}}$, $ C \gg C_{\rm{rand}}$, hierarchical means that $C(k) \sim 1/k$  and scale free refers to a power-law degree distribution with $\gamma \leq 3 $. ``TA'' and ``RA'' refer respectively to resilience  to targeted and random attacks.
}
\vspace{0.2cm}
    \centering
    \begin{tabular}{l|c|c|c}
     & Social & Myth (friendly) & Fiction  \\
  \hline
 Small world                                       & Yes       &  Yes       & Yes \\
 Hierarchy                                         & Yes       &  Yes       & Yes \\
 $p(k)$                           & Power law &  Power law & Exp.  \\
 Scale free                                        & Yes       &  Yes       & No  \\
 $G_c$                                   & $< 90\%$  &  $<90\% $  & $> 90\%$ \\
 TA                           & Vulnerable&  Vulnerable & Robust \\
 RA                           & Robust    &  Robust    & Robust \\
 Assortative                                       & Yes       &  Yes       & No
\end{tabular}
\label{tableprops}
 \end{table}

Currently there is very little evidence for the events and the society in the \emph{T\'ain}.
While there is some circumstantial evidence in terms of the landscape \cite{landscape}, its historicity is often questioned \cite{Jackson,Rahilly}.
Indeed, the social network of the full narrative initially seems similar to that of the Marvel Universe
perhaps indicating it is the Iron Age equivalent of a comic book.
However, comparing the \emph{T\'ain}'s degree distribution   to that of {\emph{Beowulf}} reveals a remarkable similarity, except for the top six vertices of the Irish narrative.
This suggests the artificiality of the network may be mainly associated with the corresponding characters.
They are similar to the superheroes of the Marvel Universe -- too super-human to be realistic, or in terms of the network, they are too well connected.
We speculate that these characters may in fact be based on amalgams of a number of entities and proxies.
To test the plausibility of this hypothesis, we removed the weak social links associated with these six characters.
The resulting network is  is assortative,   similar to the \emph{Iliad} and to other real social networks and very different to that of the Marvel Universe
and works of fiction.
We therefore suggest that if the society in the \emph{T\'ain} is to be believed, each of the top six characters is likely an amalgam that  became fused as the narrative was passed down orally through the generations.

\acknowledgments

The authors thank Christian von Ferber, Colm Connaughton and Martin Weigel for helpful comments and discussions,
Heather O'Donoghue for help with the mythological epics,
Aneesa Ayub, Rachel Findlay and Simon Rawlings for help with the data for the fictional narratives.
This work is supported by The Leverhulme Trust under grant no.~F/00~732/I.


\end{document}